\documentclass[a4paper, 12ppt]{article}
\usepackage{palatino}
\usepackage{amsfonts}
\usepackage{amssymb}
\usepackage{amsmath, amsthm}
\usepackage{mathrsfs}
\usepackage{varioref}
\usepackage{bbm}
\usepackage{picinpar}
\usepackage{graphicx}
\usepackage{algorithm, algorithmic}

\makeatletter
    \renewcommand\section{\@startsection{section}{1}{\z@}%
                                    {-3.5ex \@plus -1ex \@minus -.2ex}%
                                    {2.3ex \@plus.2ex}%
                                    {\center\bf\Large}}
    \renewcommand\subsection{\@startsection{subsection}{2}{\z@}%
                                    {-3.5ex \@plus -1ex \@minus -.2ex}%
                                    {2.3ex \@plus.2ex}%
                                    {\bf\large}}

    \newcommand\figcaption{\def\@captype{figure}\caption}
\makeatother

\renewcommand{\baselinestretch}{1.5}

\newcommand{\diff}{\,\mathrm{d}}

\newtheorem{theorem}{Theorem}
\begin{titlepage}
    \title{A probability based approach on ananlyzing dynamics of oscillators on a bidirectional ring with propagation delay}
    \author{Shuishi Yang\footnote{yang@math.miami.edu}\\School of Mathematical Sciences, Fudan University}
    \date{}
\end{titlepage}

\begin{document}
\maketitle

\begin{abstract}
    In this paper, we presented a model of pulse-coupled oscillators distributed on a bidirectional ring with propagation delay.
    In numerical simulations based on this model, we observed phenomena of
    asynchrony in a certain range of delay factor $\alpha$. To find the cause of these phenomena,
    we used a new probability based approach of analyzing. In this approach,
    the mathematical
    expectation of influence on one oscillator's phase change
    caused by its neighbor, which is regarded as a random factor, is calculated. By adding this
    expectation of influence into the firing map $h(\phi)$ introduced by Mirolla and Strogatz, a probability firing
    map $\mathbbm{H}$ is invented. By observing the behavior of $\mathbbm{H}$ 's iteration from $\mathbbm{H}$ 's graph,
    we successfully constructed a connection between the asynchrony phenomena and $\mathbbm{H}$ 's
    graph.

    \noindent\emph{keywords}: oscillator, synchronization, pulse-couple, delay, bidirectional, ring, probability
\end{abstract}

\section{Introduction \&\  Review}
    The phenomena of mutual synchronization widely exist in nature. For example, the synchronization of
    flashes of fireflies; the unison in which crickets and cicadae chirp; the pacemaker of the heart;
    and the synchronization of women's menstrual period who live together.

    A lot of researches have been done on this topic since last century. Mirolla and Strogatz\cite{MS1990} has
    studied a model of oscillators' mutual synchronization with all-to-all connection and no delay pulse
    propagation. In their model, they introduced a phase-state function $f:[0,1] \to [0, 1]$, which is required to be smooth,
    monotonic increasing, and concave down. With this $f$, the phase change of an oscillator receiving firing
    was described as $\phi^{t^{+}} = f^{-1}(f(\phi^{t^{-}} + \epsilon))$, where $\epsilon$ is the coupling strength.
    In their article, the initial conditions in wh  ich the oscillators function will not achieve synchrony,
    was proved to be of Lebesgue measure zero, given the oscillators' phase-state function satisfies the condition mentioned
    before.

    Later, the dynamics synchronization under a \emph{threshold} restriction $\phi_{c}$ was analyzed by Chia-Chu Chen \cite{Chen1993}.
    With threshold included, the phase of an oscillator will not be ``pulled up'' by others' firing when its
    phase $\phi \leqslant \phi_{c}$. Chen concluded that with $\phi_{c} < \frac{1}{2}$, synchronization can occur
    for almost all initial conditions.

    In Mathar and Mattfeldt's paper\cite{MM1996}, the same conclusion was drawn on a collection of oscillators in all-to-all
    connections, but the differentiable constraint on the phase-state function introduced in\cite{MS1990} was
    removed. Instead $f$ was only required to be strictly concave and increasing. They proved the conclusion on
    two oscillators' case by analyzing the iteration of the \emph{firing map} $h(\phi) = f^{-1}(f(1 - \phi)+\epsilon)$
    and  its fixed points. The conclusion was expanded the cases of a clique of $N$ oscillators by utilizing Fubini's law.
    In the second part of the article, a propagation delay $\alpha$ was added into the model. A similar conclusion
    that oscillators with all the initial states except a Lebesgue measure zero set will achieve synchrony was reached
    on the model with delay.

    Similar to this paper, a case of $N$ oscillators located on a unidirectional ring was discussed in \cite{VC1998}.
    In their model an oscillators pulled up to firing phase \emph{will} fire, while in Mirolla \&\ Strogatz's model\cite{MS1990} it won't.
    As a result of this mechanism, an ``avalanche'' effect exists (a chain of oscillators firing one immediately after
    another), and so helps the synchronization of the ring.

    D\'iaz-Guilera, P\'erez and Arenas\cite{DPA1998} described the \emph{FD (firing + driving)} process in the form of
    a vector's multiplying a transformation matrix $\mathbb{M}$. Thus, the evolution of states of oscillators was
    evaluated by multiplying a series of matrices on the phase vector $(\phi_{1}, \phi_{2}, \dots \phi_{N})^{T}$.
    By comparing $\mathbb{M}$'s eigenvalues with $1$, the expected evolution result of the oscillators was classified into
    converging (repeller fixed points) and phase-locking (attractor fixed points). Also the properties of
    fixed points was related with sign of $\epsilon$. Similar methods were used in \cite{GD1999}, which analyzed
    pattern formation on a ring of oscillators with couple strength $\epsilon < 0$ (denoted as $\rho$ in this paper)selection.

    In this paper, we first presented the model of oscillators distributed on a bidirectional ring with propagation delay.
    Based on this model, numerical simulation were made. From the results of the simulation, we spotted phenomena
    of asynchrony happening in a certain rang of propagation delaying factor $\alpha$ (see Fig \vref{graph:asynchrony2}).
    In purpose of explaining the asynchrony phenomena, we divided the analysis into three parts:
    \begin{description}
        \item [Step 1] For a pair of neighboring oscillators A and B, once they achieve $\alpha$-synchrony, their synchronization
        is easy to restore regardless of the influence from other neighbors. (Theorem \vref{th:restore})
        \item [Step 2] For a neighborhood consisted of oscillators B,C,D on the ring, we treat D as a random factor which uniformly
        distributed in $[0, 1)$, thus we replace D with its mathematical expectation of influence on C.
        \item [Step 3] In a four-element neighborhood (A, B, C, D), if A and B have achieved $\alpha$-synchrony once, then we
        ignore A's effect on B using the conclusion established in \emph{step 1}. Also we substitute D with its mathematical expectation
        of influence on C calculated in \emph{step 2}. After dealing with A and D, we are able to trace the synchronization process between
        B and C (See Formula \vref{BigH}).
    \end{description}
    By studying the iteration behavior of the \emph{firing map}, which describes phase change of B and C, we successfully
    made a connection between the asynchrony phenomena (Fig \vref{graph:asynchrony2}) and the graph of the firing map (Fig \vref{graphBigH}).

\section{The Model}
    In this paper we will discuss a system of $N$ oscillators distributed on a ring. This system includes the following properties:
    \begin{description}\setlength\itemsep{-.2cm}
        \item [Phase] For each oscillator $i$, there is a real number $\phi_{i} \in [0, 1]$, called a \emph{phase variable}, belongs to it to depict its phase.
        \item [Driving] As time goes by, the phase $\phi_{i}$ of each oscillator increases spontaneously at a constant speed
            $\displaystyle{\frac{\diff\phi_{i}}{\diff t}}= S = \frac{1}{T}$, here $T$ is the cycle period. For convenience
            we let $S = 1$, by which we will also have $T = 1$. See \cite{MS1990}.
        \item [Firing] When an oscillator $i$'s phase $\phi_{i}$ reaches $1$, it will fire a pulse to other oscillators immediately. (\cite{MS1990})
        \item [Ring Topology] The firing of an oscillator will be received only by those oscillators that have connection to
            the firing one. In the bidirectional ring topology of this model, $i$'s pulse is received by $j$ only if
            $j \equiv i - 1 \ \mbox{or}\ i + 1 \,(\mbox{mod} N)$. See Fig \vref{graph:cylinder1}.
        \item [Delay] Similar to \cite{MM1996}, a factor of \emph{reaction and propagation delay} is included in this model:
            an oscillator fired upon will not react to the pulse until a time delay $\alpha$ passes, (i.e. $i$ fires at $t_{0}$, $j$ only reacts at $t_{0} + \alpha$).
        \item [Period of no responding]
            The oscillator will not react if its phase is in the interval $[0, 2\alpha)$
            i.e. $j$ will not react if $\phi_{j} \in [0, 2\alpha)$ at $t_{0} + \alpha$.
        \item [Coupling] When an oscillator $j$ reacts to a pulse (with precondition, $\phi_{j}^{t^{-}} \notin [0, 2\alpha) \mbox{ at } t$ as
            the pulse is fired at $t - \alpha$), its phase is ``pulled forward'' for $\rho\cdot\phi_{j}^{t^{+}} + \epsilon$. That is
            $$
                \phi_{j}^{t^{+}} =
                \begin{cases}
                    (1 + \rho)\cdot \phi_{j}^{t^{-}} + \epsilon \quad\quad& \mbox{if }(1 + \rho)\cdot \phi_{j}^{t^{-}} +\epsilon < 1 \\
                    0 & \mbox{if }(1 + \rho)\cdot \phi_{j}^{t^{-}} + \epsilon \geqslant 1
                \end{cases}
            $$
            Here $\rho$ denotes the coupling strength, and $\epsilon$ the coupling constant. In this model we always have $\rho > 0, \epsilon > 0$.
            Note that when an oscillator's $\phi$ is ``pulled''
            up by others to reach 1, it \emph{will} still fire.
        \item [Phase difference] The phase difference between two oscillators $i$ and $j$ are defined as
            $$
                \mathbbm{D}(\phi_{i}, \phi_{j}) =
                \begin{cases}
                    |\phi_{i} - \phi_{j}| \quad\quad &\mbox{if } |\phi_{i} - \phi_{j}| \leqslant \frac{1}{2} \\
                    1 - |\phi_{i} - \phi_{j}| &\mbox{if } |\phi_{i} - \phi_{j}| > \frac{1}{2}
                \end{cases}
            $$
            With this definition, the difference between $\phi_{i} = 0.9 $ and $\phi_{j} = 0.1$ is $0.2$, not $0.8$.
        \item [$\alpha$-synchronization]
            When oscillators $i$ and $j$ satisfy $\mathbbm{D}(\phi_{i}, \phi{j}) \leqslant \alpha$, they are
            \emph{$\alpha$-synchronized}, or in other words, achieved \emph{$\alpha$-synchronization}. See \cite{MM1996}.
            In addition, in a system of oscillators distributed on a ring, this system achieves $\alpha$-synchronization
            when each pair of ``neighbor'' achieves \quad $\alpha$-synchronization: $\forall i$, if $j \equiv i - 1 \ \mbox{or}\ i + 1 \,(\mbox{mod }N)$ $\Longrightarrow$
            $\mathbbm{D}(\phi_{i}, \phi_{j}) \leqslant \alpha $. A system that has achieved $\alpha$-synchronization
            is showed in Fig \vref{graph:cylindersyn}.
        \item [Respond only once] An oscillator responds to a pulse from the same ``neighbor'' only once in one firing cycle.
            e.g. If in some way $i$ ``catches'' up with $j$ and fires the second time before $j$'s firing, $j$ does not respond to $i$'s second pulse.
            So far no previous models has added this restriction.
    \end{description}
\section{Numerical simulation}
    Before presenting a theoretical analysis on the dynamics of the system, first we provide some results
    on the numerical simulation.

    \subsection{Process of synchronization}
\begin{figure}
            \begin{minipage}[t]{8cm}
                \includegraphics[width = 4cm]{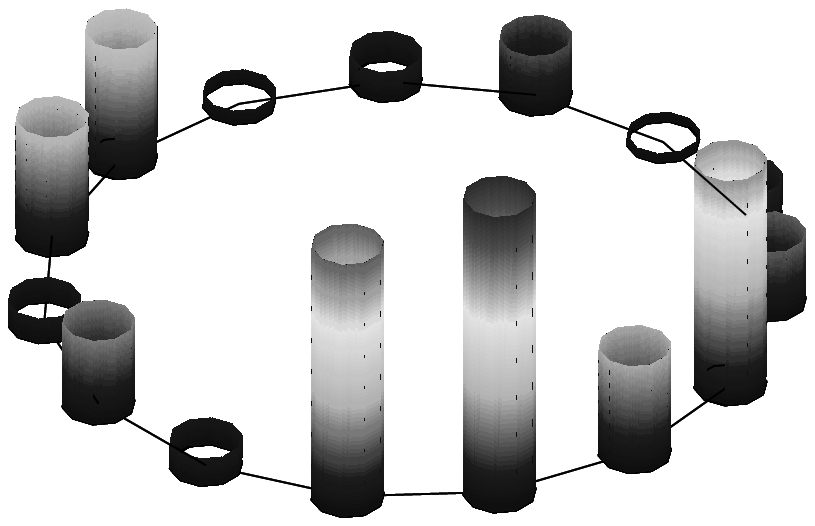}
                \caption{
                    \label{graph:cylinder1}\small
                    A ring of oscillators represented by cylinders. The height of cylinder depicts the phase of an oscillator.
                    The connection lines on the bottom indicate the neighboring relationship.
                }
            \end{minipage}
            \begin{minipage}[t]{8cm}
                \includegraphics[width = 4cm]{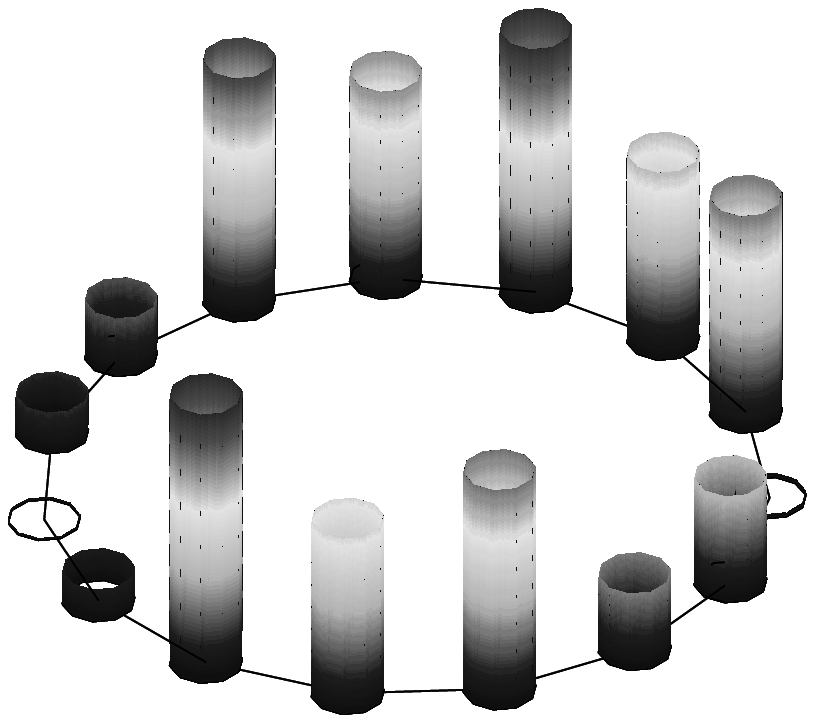}
                \caption{
                    \label{graph:cylinder2}\small
                    A system in its synchronization process.
                }
            \end{minipage}
            \begin{minipage}[t]{9cm}
                \includegraphics[width = 4cm]{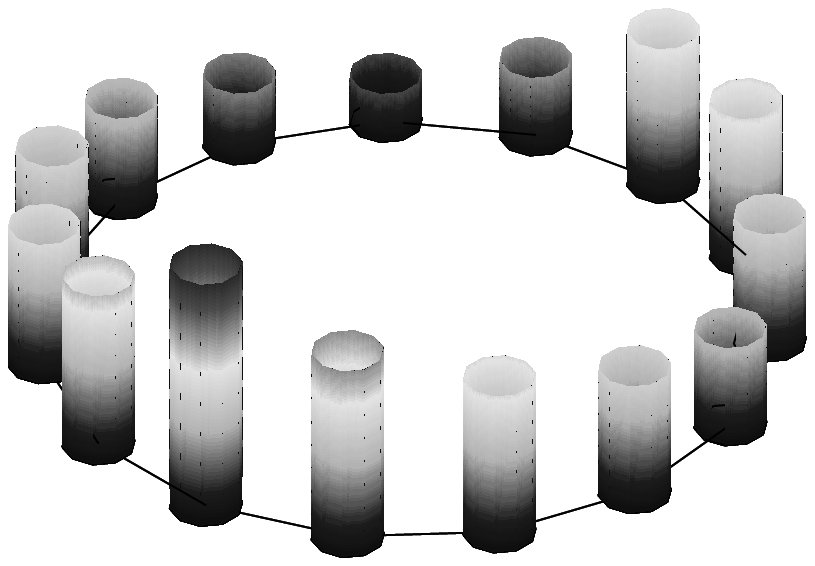}
                \caption{
                    \label{graph:cylindersyn}\small
                    A system that achieved $\alpha$-synchronization
                }
            \end{minipage}
        \end{figure}

        The first experiment tracks the process of synchronization. In purpose of inspecting
        the course of a system's synchronizing, a demonstration program is written in MATLAB, which
        shows the dynamics of oscillators represented by cylinders. Fig \ref{graph:cylinder1}, \ref{graph:cylinder2},
        \ref{graph:cylinder3} and \ref{graph:cylindersyn} display four snapshots in a synchronization procedure.

\begin{figure}[htbp]
            \includegraphics[width = 8cm]{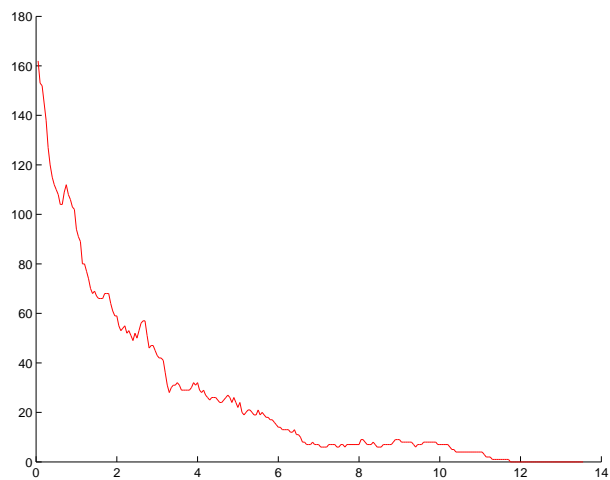}
            \caption {
                \label{graph:asyncount}
                Count of pairs in asynchrony with the change of time. The system parameters are
                $\alpha = 0.1$, $\rho = 0.3$, $\epsilon = 0.01$, $N = 200$
            }
        \end{figure}

        Also the count of $\alpha$-synchronized pairs ( $(\phi_i, \phi_{i + 1})$ s.t. $\mathbbm{D}(\phi_i, \phi_{i + 1}) \leqslant \alpha$)
        is tracked in a numerical experiment. In Fig \ref{graph:asyncount},
        the $x$ axis represents time, while $y$ axis represents the count of neighbors that are not
        $\alpha$-synchronized. From this figure we see that the trend of this system is achieving
        $\alpha$-synchrony.
        In addition, a figure similar to \cite{VC1998} 's grey level denotation is made. In Fig \vref{graph:greylevel},
        $x$ axis is the index of oscillators (that is to say, two pixels horizontally adjoining each other are neighboring
        oscillators $i$ and $i + 1$), $y$ axis represents time (time increases via the direction of up-down), and the
        greylevel of a pixel $(i, t)$ on this figure depicts the phase $\phi_i$ of oscillator $i$ at time $t$ (the whiter,
        the nearer to 1, the blackest represents $\phi = 0$). This figure is a vivid record of the whole
        synchronization process.
    \subsection{Testing time cost to achieve $\alpha$-synchronization}
\begin{figure}
            \begin{minipage}[c]{12cm}
                \includegraphics[width = 8cm]{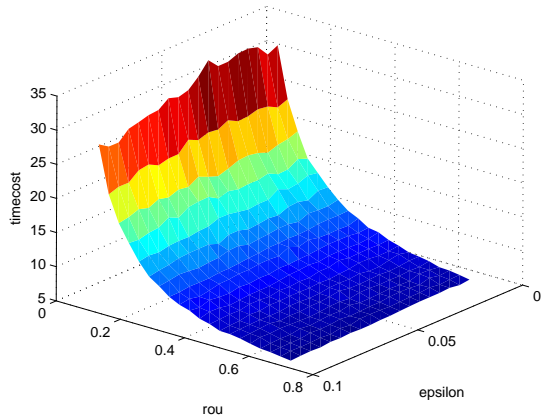}
            \end{minipage}
            \begin{minipage}[c]{8cm}
                \caption {
                    \label{graph:rho-epsilon-t1}
                    Average time cost to achieve $\alpha$-synchronization in 100 systems with different $\rho$ and $\epsilon$.
                }
            \end{minipage}
        \end{figure}

        An experiment comparing the time used to achieve $\alpha$-synchronization with different system
        parameters is conducted. In the simulation, we tested systems with $N$ oscillators, and repeated
        the test for 100 times on 100 randomly generated samples, for each parameter combination.
        We generated 100 initial condition samples $\{S_1, S_2, \dots, S_{100}\}$, each
        $S_k = \{\phi_0^{(k)}, \phi_1^{(k)}, \dots, \phi_{N-1}^{(k)}\}$. For each combination of parameters such
        as $\alpha$, $\rho$, $\epsilon$, an average time used to achive $\alpha$-synchronization of
        the 100 samples is calculated. See algorithm (\ref{alg:timecost}) for a more strict description of the algorithm.
        \begin{algorithm}
            \caption{Calculating the time cost to achieve $\alpha$-synchronization with differnt $\rho$ and $\epsilon$}
            \label{alg:timecost}
            \begin{algorithmic}
                \STATE Randomly generate 100 initial samples $\{S_1, S_2, \dots, S_{100}\}$.
                \STATE Generate the set of parameter values: $\widetilde{\rho} = \{\rho_1, \rho_2, \dots, \rho_l\}$, $\widetilde{\epsilon} = \{\epsilon_1, \epsilon_2, \dots, \epsilon_m\}$.
                \FORALL {$\rho \in \widetilde{\rho}$}
                    \FORALL {$\epsilon \in \widetilde{\epsilon}$}
                        \STATE $\text{timecost} \leftarrow 0$;
                        \FORALL {$1 \leqslant i \leqslant 100$}
                            \STATE timecost += (the time used to achieve $\alpha$-synchronization by $S_i$ under $\rho$, $\epsilon$)
                        \ENDFOR
                        \STATE Draw a point $(\rho, \epsilon, \displaystyle\frac{\text{timecost}}{100})$on the $\rho-\epsilon-t$ graph.
                    \ENDFOR
                \ENDFOR
            \end{algorithmic}
        \end{algorithm}
        A result is showed in Fig \vref{graph:rho-epsilon-t1}.
    \subsection{Phenomena of asynchrony}
\begin{figure}[hbpt]
            \begin{minipage}[t]{7cm}
                \includegraphics[width = 8cm]{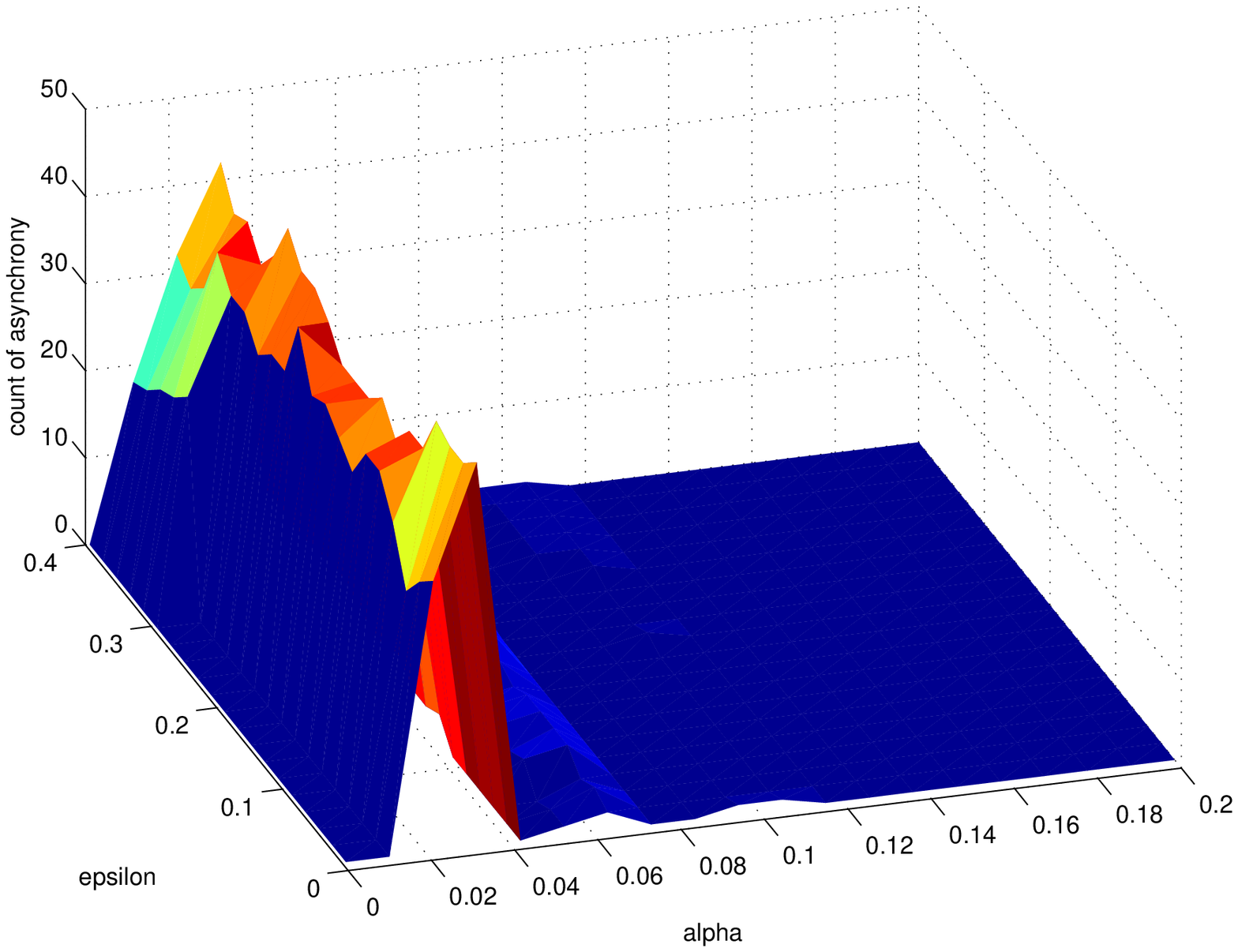}
                \caption {
                    \label{graph:asynchrony1}
                    A $\alpha-\epsilon-$count graph that shows the count of configurations which don't achieve synchrony before $t = 100$.
                    $\rho = 0.3$, $N = 200$, out of 100 samples.
                }
            \end{minipage}
            \hfill
            \begin{minipage}[t]{7cm}
                \includegraphics[width = 8cm]{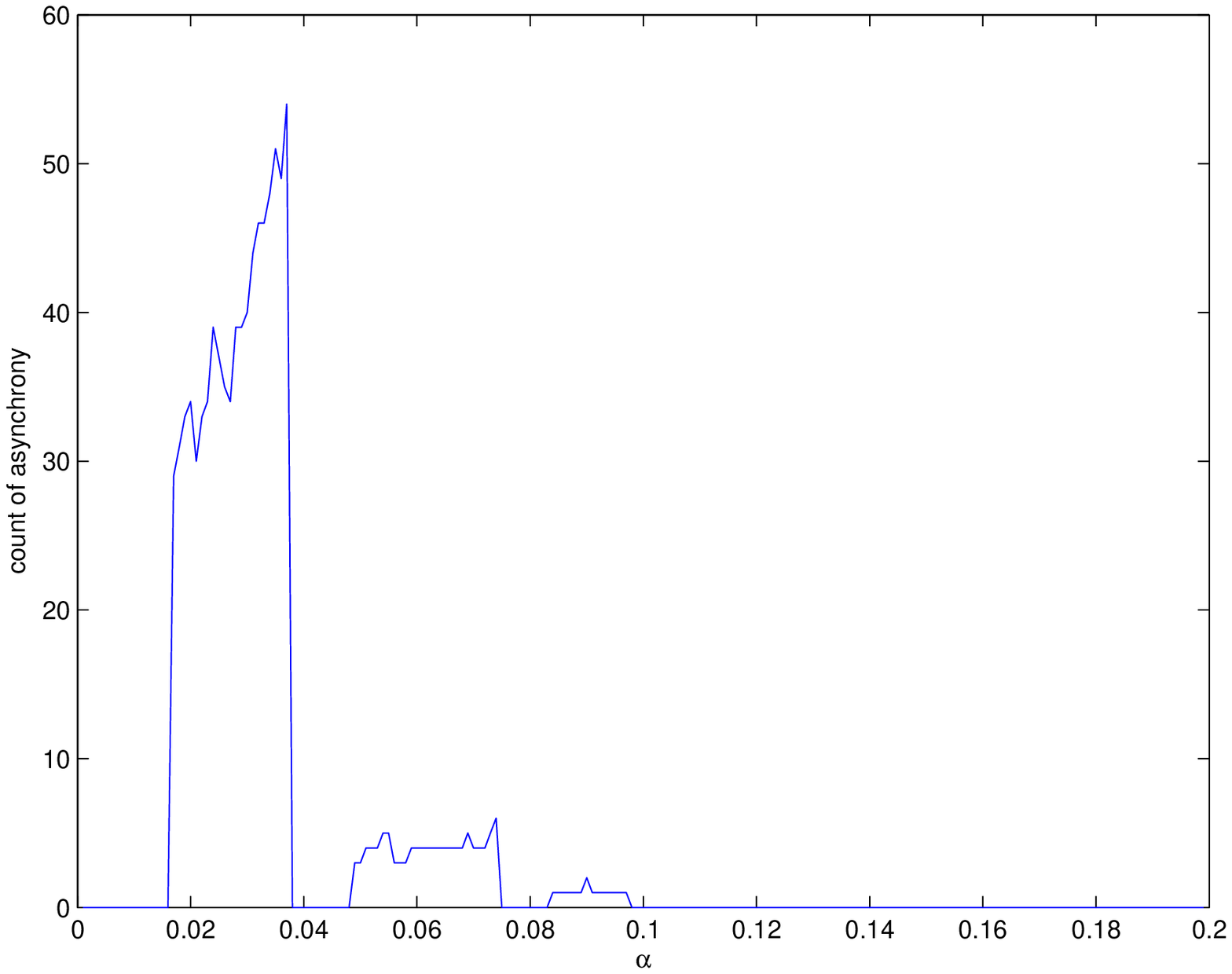}
                \caption {
                    \label{graph:asynchrony2}
                    Count of initial configurations that fail to achieve $\alpha$-synchronization before $t = 100$, out of
                    100 samples. $\rho = 0.3$, $\epsilon = 0.01$, $N = 200$.
                }
            \end{minipage}
        \end{figure}

        It is not always that, with any parameter, the system can eventually achieve $\alpha$-synchronization.
        Similar to algorithm \ref{alg:timecost}, a program to count the configurations that fail to
        achieve $\alpha$-synchronization with a certain combination of parameters is designed.
        From Fig \ref{graph:asynchrony1}, it is easy to see that with $\alpha \in [0.01, 0.1]$, asynchrony frequently
        happens.

        For a clearer observation, a 2-D plot with respect to $\alpha-$count is drawn on Fig \ref{graph:asynchrony2}.
        Obviously, with $\alpha \approx 0.02$ and $\alpha \approx 0.05$, the chance of an asynchrony is large, while
        with $\alpha \geqslant 0.1$ the experiment always end in synchrony. In purpose of trying to find a reason
        to these phenomena, an analysis of the whole section \ref{section:dynamics} is implemented.

 \section{Analysis of the Dynamics}
    \label{section:dynamics}
    In purpose of explaining phenomena observed in Fig \vref{graph:asynchrony2}, a step by step analysis is made in this
    section. First, we will review some useful notions introduced in previous works; then a new probability-based
    approach will be presented to find a cause of Fig \vref{graph:asynchrony2}.
    \subsection{Firing map}
        Similar to \cite{MM1996}, first we check the dynamics of a system of two oscillators: A, B. Now Let a row vector
        $(\phi_{1}, \phi_{2})$ to represent the state of two oscillators. Without loss of generality, we can always
        ``shift'' the phases of A, B into a form of $(\alpha, \phi)$, note that in this representation we change
        the order of components to make sure $\phi > \alpha$. After a FD process, the state of this pair becomes
        $(\alpha, \phi')$ (a change of order from $(\phi', \alpha)$). We define function $h:[0,1)\to[0,1)$ to describe the mapping $\phi\mapsto\phi'$. Without
        considering the bounds, $h$ may simply be
        \begin{equation}
            h(\phi) = (1 + \rho)\cdot(1 + \alpha - \phi) + \epsilon
        \end{equation}
        Now consider the restrictions on $h(\phi)$:
        \begin{enumerate}
            \item when $\phi \in [0, \alpha)$, there's no definition on $h$ since we suppose $\phi \geqslant \alpha$.
                But for convenience we expand the domain of $h$ to $[0, 1)$
            \item when $\phi \in [\alpha, 2\alpha)$, since they achieve $\alpha$-synchronization,
                there's no effect on A from B's firing.
            \item when $\phi$ and $\alpha$ are ``farther'' away from $\alpha$, but still too ``close'', B's firing
                will pull A's phase to reach $1$ and cause A's immediate firing.
                We can calculate this bound $\delta_{\alpha}$:\\
                $h(\delta_{\alpha}) = 1 \Longleftrightarrow (1 + \rho)\cdot(1 + \alpha - \delta_{\alpha}) + \epsilon = 1
                \Longleftrightarrow$
                \begin{equation}
                    \delta_{\alpha} = 1 + \alpha - \displaystyle{\frac{1 - \epsilon}{1 + \rho}}
                \end{equation}
        \end{enumerate}
        Based on the discussion above, we define the firing map
        \begin{equation}
            h(\phi) =
            \begin{cases}
                2\alpha - \phi \quad & \mbox{if } 0 \leqslant \phi < 2\alpha \\
                0 & \mbox{if } 2\alpha \leqslant \phi \leqslant \delta_{\alpha} \\
                (1 + \rho)\cdot(1 + \alpha - \phi) + \epsilon & \mbox{if } \delta_{\alpha} < \phi < 1
            \end{cases}
            \label{def:h}
        \end{equation}
    \subsection{Jumping length}
        To inspect A's ``length'' of phase that was ``pulled'' forward by B's firing, we introduce
        the concept of \emph{Jumping Length}. When A responds to B's firing at time $t$,
        A's jumping length is $\mathbbm{D}(\phi_{\text{A}}^{t^{+}}, \phi_{\text{A}}^{t^{-}})$. Furthermore, on
        a standard two-oscillator configuration $(\alpha, \phi)$, the jumping length that A \emph{will} have
        due to B's firing is defined as
        \begin{equation}
            \mathcal{J}(\phi) =
            \begin{cases}
                0 \quad &\mbox{if } \phi \in [0, 2\alpha)\\
                \phi - 2\alpha  & \mbox{if } \phi \in[2\alpha, \delta_{\alpha}]\\
                \rho\cdot(1 + \alpha - \phi) + \epsilon & \mbox{if } \phi \in (\delta_{\alpha}, 1)
            \end{cases}
            \label{def:J}
        \end{equation}
        And for a non-standard configuration $(\phi_{\text{A}}, \phi_{\text{B}})$, where $(\phi_{\text{A}} < \phi_{\text{B}})$,
        the jumping length of A that will be caused by B:
        $\mathcal{J}^{\ast}\left(\phi_{\text{A}}\leftarrow\phi_{\text{B}}\right)$
        is defined as
        \begin{equation}
            \mathcal{J}^{\ast}\left(\phi_{\text{A}}\leftarrow\phi_{\text{B}}\right)
            = \mathcal{J}\left(\phi_{\text{B}} - \phi_{\text{A}} + \alpha\right) \label{def:JStar}
        \end{equation}

    \subsection{Stability of $\alpha$-synchronization}
        In a system of two oscillators, once the system achieves $\alpha$-synchronization, the synchrony will never be lost
        as the oscillators don't respond to each other's pulse all the time regarding to the rule \emph{delay} in the model.
        But in the system of more than $2$ oscillators, each one is influenced by its two neighbors.
        Will a pair of oscillators that achieves $\alpha$-synchronization lose synchrony due to others' coupling?
        Here we can show that if a pair is $\alpha$-synchronized once, they will restore their synchrony
        any time one of them reacts to the other one's firing.
        \begin{theorem}
            $\phi_i$ and $\phi_{i + 1}$ ($\phi_i)$
            are neighboring oscillators that achieve $\alpha$-synchronization at one time: \\
            $\exists t_{0}: \mathbbm{D}(\phi_i(t_{0}), \phi_{i+1}(t_{0})) \leqslant \alpha$.
            Then for any $t^{\ast} > t_{0}$ at which  $i$ or $i + 1$ fires, we will have\\
            $\mathbbm{D}(\phi_i(t^{\ast}+\alpha), \phi_{i+1}(t^{\ast}+\alpha)) \leqslant \alpha$ again.
            \label{th:restore}
        \end{theorem}
        \begin{proof}
            Without loss of generality, we suppose $\phi_{i + 1} = \phi_{i} + d$, $0 \leqslant d \leqslant \alpha$, and
            $k \in \{ i , i + 1\}$ the oscillator that will first fire (note that $i$ may ``catch up'' with $i + 1$).
            Let $t^{\ast}$ be the time of $k$'s firing.
            \begin{description}
                \item [Case 1 --- Neither $i$ or $i + 1$ receives any pulse in [$t_0 - \alpha, t^{\ast}$\expandafter{]} :]{\quad}\\
                    \begin{minipage}{8cm}
                        Obviously without the influence of pulse outside the pair $i$ and $i + 1$, these two oscillators will
                        remain in synchrony:
                        $$
                            \mathbbm{D}(\phi_i(t)), \phi_{i+1}(t)) \leqslant \alpha, \forall t \in [t_{0}, t^{\ast} + \alpha]
                        $$
                        (See the figure on the right: the height of bars represents phase $\phi$, the two bars in the middle
                        are $i$ and $i+1$. The leftmost and rightmost bars will not fire before $i + 1$'s firing.)
                    \end{minipage}
                    \hfill
                    \begin{minipage}{8cm}
                        \includegraphics[width = 8cm]{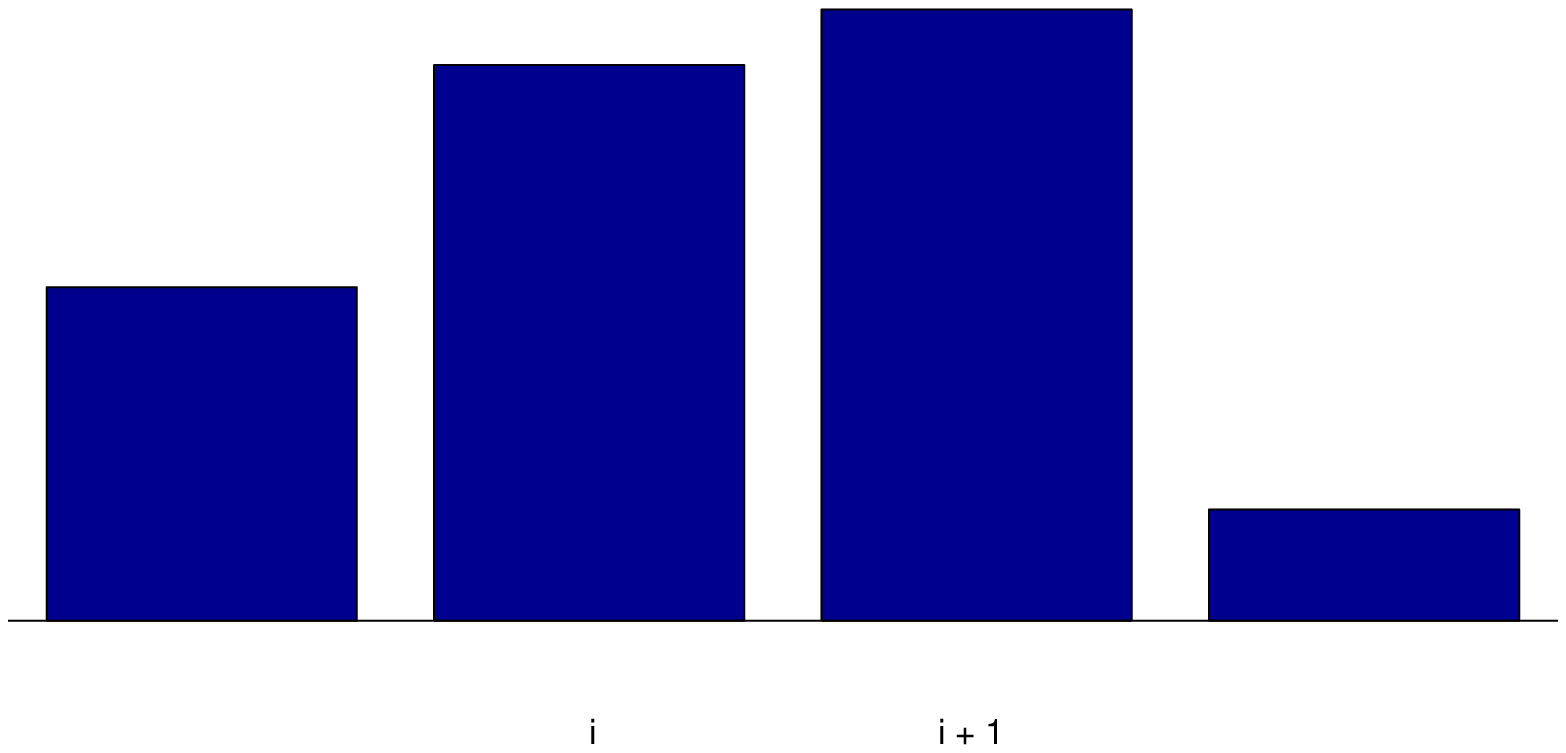}
                    \end{minipage}
                \item [Case 2 --- $i + 1$ receives a pulse and $i$ doesn't receive one before $t^{\ast}$:]{\quad}\\
                Suppose $i + 2$ fires at $t_1$, then $i + 1$ reacts to its pulse at $t_2 = t_1 + \alpha$
                    \begin{description}
                        \item [(i)$(1 + \rho)\cdot\phi_{i+1}^{t_2^-} + \epsilon \geqslant 1$:] \quad \\
                        As the phase of $i + 1$ reaches $1$, $i + 1$ fires immediately and resets to 0 (by the way we have $t^{\ast} = t_2$)\\
                        After a time length of $\alpha$, $i$ reacts to the firing at $t_3 = t^{\ast} + \alpha$.\\
                        $\because \phi_{i + 1} = \phi_{i} + d$, $d \leqslant \alpha$,\\
                        $\therefore \phi_{i}^{t_3^{-}} = \phi_{i}^{t^\ast} + \alpha = \phi_{i + 1}^{t^{\ast-}} - d + \alpha \geqslant \phi_{i + 1}^{t^{\ast-}}$\\
                        $
                            \Longrightarrow \phi_{i}^{t_3^{+}} = (1 + \rho)\cdot\phi_{i}^{t_3^{-}} + \epsilon \geqslant (1 + \rho)\cdot\phi_{i + 1}^{t^{\ast-}} + \epsilon
                            = (1 + \rho)\cdot\phi_{i + 1}^{t_2^{-}} + \epsilon \geqslant 1
                        $\\
                        That is to say, $\phi_{i}$ is reset to $0$ at $t^{\ast} + \alpha$,\\
                        $\Longrightarrow \mathbbm{D}(\phi_{i}, \phi_{i + 1}) = \mathbbm{D}(0, \alpha) = \alpha$,\\
                        which indicates $i$ and $i + 1$ are $\alpha$-synchronized again.
                        \item [(ii)$(1 + \rho)\cdot\phi_{i+1}^{t_2^-} + \epsilon < 1$:] \quad \\
                            After a time length $l = 1 + \alpha - \phi_{i + 1}^{t_2^+}$, $i$ will respond to $i+1$'s firing.\\
                            Let $t_3 = t^{\ast} + \alpha$ ( also $= t2 + l$), then
                            \begin{eqnarray}
                                (1 + \rho)\cdot\phi_{i}^{t_3^-} + \epsilon
                                                &=& (1 + \rho)\cdot(\phi_{i}^{t_2} + l) + \epsilon \nonumber\\
                                                &=& (1 + \rho)\cdot(\phi_{i + 1}^{t_2^-} - d + \alpha + 1 - \phi_{i + 1}^{t_2^+}) + \epsilon \nonumber\\
                                                &>& (1 + \rho)\cdot(\phi_{i + 1}^{t_2^-}) + \epsilon + (1 - \phi_{i + 1}^{t_2^+})  \label{ineq:d}\\
                                                &=& \phi_{i + 1}^{t_2^+} + (1 - \phi_{i + 1}^{t_2^+}) \nonumber\\
                                                &=& 1 \nonumber
                            \end{eqnarray}
                            Hence $\phi_{i}$ is reset to 0 at $t_3^+$, which implies
                            $\mathbbm{D}(\phi_{i}, \phi_{i + 1}) = \mathbbm{D}(0, \alpha) = \alpha$ after $i$ responds to the pulse from $i+1$\\
                            $\Longrightarrow$ the $\alpha$-synchronization remains.
                    \end{description}
                    So in case 2, the $\alpha$-synchronization will restore.
                \item [Case 3 --- $i$ receives a fire before $t^\ast$]\ \\
                    This case is divided into two sub-cases with respect to $k$ (whether $i$ ``catches up'' with $i + 1$ or not).
                    \begin{description}
                        \item [(i) $k = i + 1$] \ \\
                            $i + 1$ will be the first that fires and $i$ will reacts to its firing at $t_3 = t^\ast + \alpha$\\
                            Imagine there is a ``shadow'' oscillator $\widetilde{i}$, which has a phase equal to $i$ at $t_0$: $\widetilde{\phi_{i}}^{t_0} = \phi_{i}^{t_0}$,
                            but $\widetilde{i}$ doesn't respond to $i - 1$'s firing during $[t_0, t^\ast + \alpha]$.\\
                            Then $\widetilde{i}$ 's case is discussed in \textbf{case 2}.\\
                            $\Longrightarrow$ $\widetilde{\phi_{i}}^{t_3^{+}} \geqslant 1$.\\
                            And obviously we have $\widetilde{\phi_{i}}^{t_3^{-}} < \phi_{i}^{t_3^{-}}$ \\
                            $\Longrightarrow$ $\phi_{i}^{t_3^{+}} \geqslant \widetilde{\phi_{i}}^{t_3^{+}} \geqslant 1$,\\
                            which is a proof a $i$ and $i+1$'s restoration of $\alpha$-synchronization.
                        \item [(ii) $k = i$]\ \\
                            If $i + 1$ doesn't receive a fire from $i + 2$ before $t_3 = t^\ast + \alpha$,\\
                            Similar to \textbf{Case 2(ii)}, suppose $i - 1$ fires at $t_1$, then $i$ reacts to its pulse at $t_2 = t_1 + \alpha$\\
                            After a time length $l = 1 + \alpha - \phi_{i}^{t_2^+}$, $i + 1$ will respond to $i$'s firing.\\
                            No matter $i + 1$ reacts to $i + 2$ 's pulse before $t^\ast$ or not, we have
                            \begin{eqnarray}
                                (1 + \rho)\cdot\phi_{i + 1}^{t_3^-} + \epsilon
                                                &\geqslant& (1 + \rho)\cdot(\phi_{i + 1}^{t_2} + l) + \epsilon \nonumber\\
                                                &\geqslant& (1 + \rho)\cdot(\phi_{i}^{t_2^-} + d + \alpha + 1 - \phi_{i}^{t_2^+}) + \epsilon \nonumber\\
                                                &>& (1 + \rho)\cdot(\phi_{i}^{t_2^-}) + \epsilon + (1 - \phi_{i}^{t_2^+})  \nonumber\\
                                                &=& \phi_{i}^{t_2^+} + (1 - \phi_{i}^{t_2^+}) \nonumber\\
                                                &=& 1 \nonumber
                            \end{eqnarray}
                            That is $\mathbb{D}(\phi_{i}^{t_3}, \phi_{i + 1}^{t_3^+}) = \mathbb{D}(\alpha, 0) = \alpha$.
                    \end{description}
                    $\therefore$ the $\alpha$-synchronization remains if $i$ receives a fire before $t^\ast$, no matter $i+1$ receives or not.
            \end{description}
            So we have proven that an $\alpha$-synchronization will restore at the time when one of the once-synchronized oscillators reacts to its synchronized partner's pulse.
        \end{proof}
    \subsection{Estimation on Dynamics}
        The dynamics of a system on a bidirectional ring with delay is complex: each oscillator in it responds to both side
        of neighbors, while each neighbor is influenced by a farther neighbor. It's not easy to trace the evolution
        of the system based on an exact event-tracing approach like in \cite{MS1990}, or \cite{VC1998} as the order of firing
        will not change due to unidirectional connection and non-delay in their cases. So, in this model,
        we need to use a new probability based approach to study the dynamics of the system.

        \subsubsection{Neighborhood of three}
            Now consider a  three-oscillators-composed neighborhood (B, C, D), in which B and C have not achieved $\alpha$-synchronization
            (because the purpose of this analysis is on the synchronization process of B and C).
            Let a standard configuration $(\alpha, \phi, \hat\phi)$, ($\phi \geqslant 2\alpha$)be used to describe the starting state of this neighborhood at $t_0$.
            For a given $\hat\phi$, if $\hat\phi$ is in some certain area(which will be discussed in the next paragraph), C will
            have a coupling effect on B in current cycle.
            That is after a time of $(1 + \alpha - \hat\phi)$, $\phi$ is added up with
            $\mathcal{J}^{\ast}(\phi_\leftarrow\hat{\phi})$ --- i.e. the state becomes
            $(\alpha + (1 + \alpha - \hat\phi), \phi + (1 + \alpha - \hat\phi) +
            \mathcal{J}^{\ast}(\phi_\leftarrow\hat{\phi}), \alpha)$
            at $t_0 + (1 + \alpha - \hat\phi)$.
        \subsubsection{Domain of $\hat\phi$}
            For a given $\phi$, there is an interval $\mathcal{D}_\phi$ which $\hat\phi$ must be in, in order to affect $\phi$ in C's current cycle.
            $\mathcal{D}_\phi$ is defined piecewise:
            $$
                \mathcal{D}_\phi =
                \begin{cases}
                    (\phi + \alpha, 1) \cup [0, \alpha) \quad &\text{ if } \phi \in (2\alpha, 1 - \alpha)\\
                    (\alpha + \phi - 1 , \alpha) & \text{ if } \phi \in (1 - \alpha, 1)
                \end{cases}
            $$
            To unite the definition, we extend $\hat\phi$ 's domain from $[0, 1)$ and treat $1 + \hat\phi$ the same as $\hat\phi$.
            In this way $\mathcal{D}_\phi$ becomes
            $$
                \mathcal{D}_\phi = (\phi + \alpha, 1 + \alpha)
            $$

        \subsubsection{Reducing three to two}
            With respect to D's influence on C, there is no difference whether C's ``adding up'' $\mathcal{J}^{\ast}(\phi_\leftarrow\hat{\phi})$
            happens after a while or immediately.
            So, in purpose of analyzing the dynamics of (B, C) in one cycle, we can treat D's coupling as if it happens immediately at $t_0$.
            That is, for this neighborhood $(\alpha, \phi, \hat\phi)$, we can depict its state with one less opponent:
            $(\alpha, \phi + \mathcal{J}^{\ast}(\phi_\leftarrow\hat{\phi}))$
            ,  by making D's pulse into effect immediately and discard it from this neighborhood for its effect in this cycle is already done.
            So the representation of this neighborhood $(\alpha, \phi, \hat\phi)$ becomes $(\alpha, \phi + \mathcal{J}^{\ast}(\phi_\leftarrow\hat{\phi}))$.

        \subsubsection{Random factor}
            By putting the neighborhood (B, C, D) back to the ring of oscillators, we may notice that
            B is also influenced by its left neighbor A,
            while D is affected by its right neighbor E. If we go on inspecting A and E,
            we have to face a very long chain of cause-and-effects.
            So, now we use a probability based approach to analyze the possible behavior of the system. That is, when analyzing
            the behavior of B and C, we consider A and D as random factors --- their phases uniformly distribute on $[0, 1)$.

        \subsubsection{Expectation of influence}
            For a standard configuration of B, C: $(\alpha, \phi)$, C is possible to be influenced by D's firing before its reaching $0$.
            So we consider D's phase $\hat{\phi}$ a random variable uniformly distributed on $[0, 1)$. When $\hat{\phi} \in \mathcal{D}_\phi$,
            C will react to D's firing before its own resetting. This response causes C's phase to have a jumping forward of length
            $\mathcal{J}^{\ast}(\phi_\leftarrow\hat{\phi})$. So the mathematical expectation of C's jumping length caused by
            random variable $\hat{\phi}$ is
            \begin{eqnarray}
                & & E(\mathcal{J}^{\ast}(\phi_\leftarrow\hat{\phi}))\nonumber\\
                &=& P\{\hat{\phi} \in \mathcal{D}_\phi\}\cdot\frac{1}{\parallel\mathcal{D}_\phi\parallel}\int_{\mathcal{D}_\phi}\mathcal{J}^{\ast}(\phi_\leftarrow\hat{\phi}) \,\mathrm{d}\hat{\phi}\nonumber\\
                &=& \frac{1 - \phi}{1 - \phi}\int_{\mathcal{D}_\phi}\mathcal{J}^{\ast}(\phi_\leftarrow\hat{\phi}) \,\mathrm{d}\hat{\phi} \nonumber\\
                &=& \int_{\mathcal{D}_\phi}\mathcal{J}^{\ast}(\phi_\leftarrow\hat{\phi}) \,\mathrm{d}\hat{\phi}\label{EJump}
            \end{eqnarray}

\begin{figure}
                \includegraphics[width = 8cm]{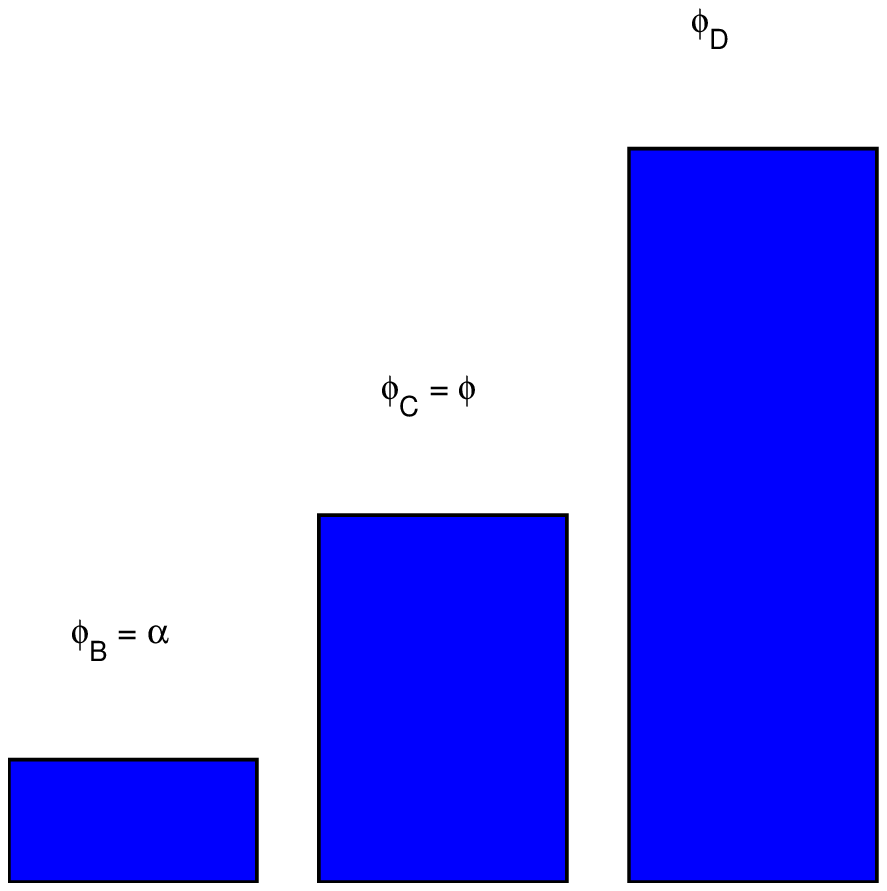}
                \includegraphics[width = 8cm]{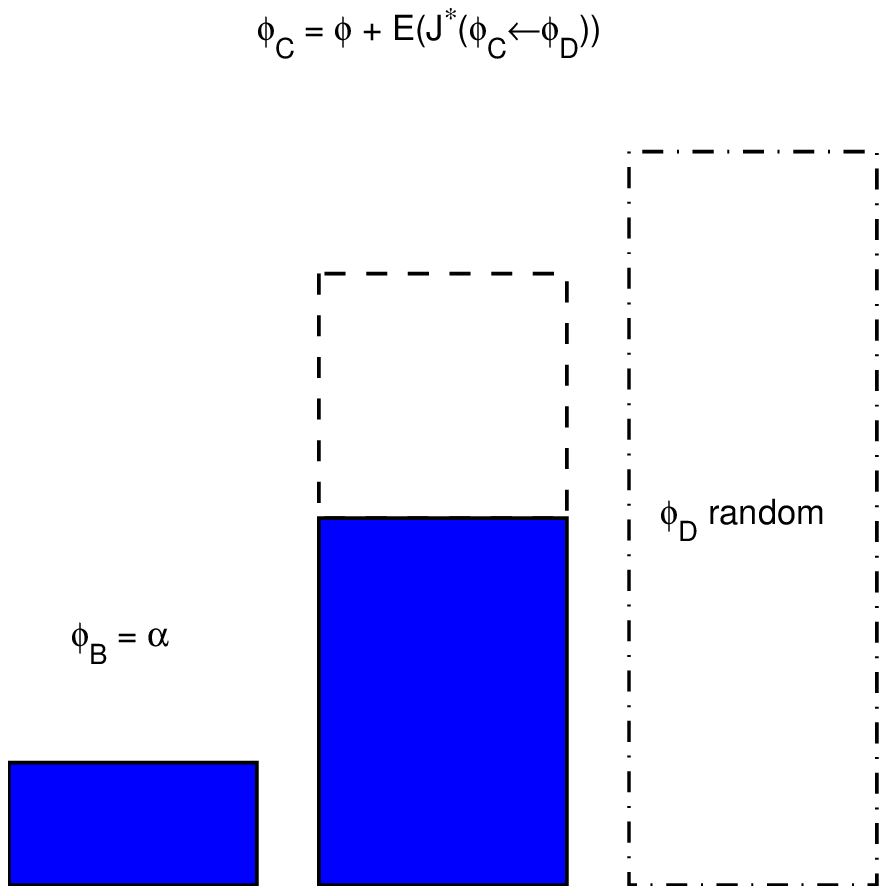}
                \caption{
                    \label{graph:randomreplace} A demonstration of replacing $\phi_{\text{D}}$ with the mathematical
                    expectation of influence that it may cause on C.
                }
            \end{figure}

            So in the neighborhood consisted by B, C, D ($\alpha, \phi, \hat\phi$), we regard D as a random factor and replace
            it by its ``average'' affection $E(\mathcal{J}^{\ast}(\phi_\leftarrow\hat{\phi}))$ on C. Thus
            we can analyze B,C's dynamics in a cycle by inspecting $(\alpha, \phi + E(\mathcal{J}^{\ast}(\phi_\leftarrow\hat{\phi})))$.
            This replacement is demonstrated in Fig \ref{graph:randomreplace}.

\begin{figure}[htbp]
                \begin{minipage}[t]{8cm}
                    \includegraphics[width = 8cm]{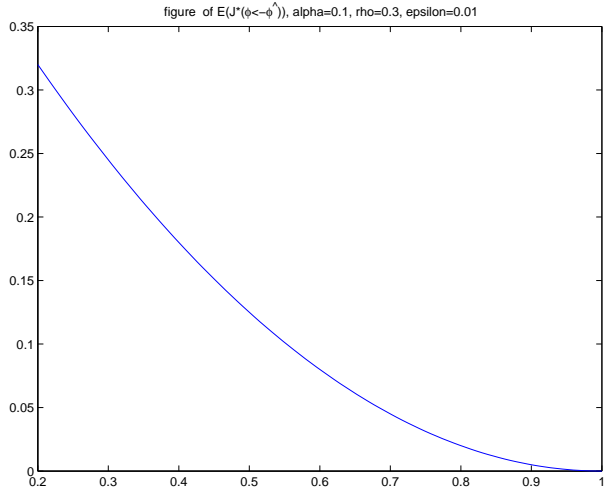}
                \end{minipage}
                \begin{minipage}[b]{3cm}
                    \caption {
                        \label{graph:EJ}
                        A plot of $E(\mathcal{J}^\ast(\phi\leftarrow\hat\phi))$
                    }
                \end{minipage}
            \end{figure}

            To calculate $\phi + E(\mathcal{J}^{\ast}(\phi_\leftarrow\hat{\phi}))$ using (\ref{EJump}) and (\ref{def:JStar}), we have
            \begin{eqnarray}
                & & \phi + E(\mathcal{J}^{\ast}(\phi_\leftarrow\hat{\phi}))\nonumber\\
                &=& \phi + \int_{\mathcal{D}_\phi}\mathcal{J}^{\ast}(\phi_\leftarrow\hat{\phi}) \,\mathrm{d}\hat{\phi}\nonumber\\
                &=& \phi + \int_{\phi + \alpha}^{1 + \alpha}\mathcal{J}(\hat\phi - \phi + \alpha) \,\mathrm{d}\hat\phi\nonumber\\
                &=& \phi + \int_{2\alpha}^{1 + 2\alpha - \phi}\mathcal{J}(s) \,\mathrm{d}s\nonumber\\
                &=&
                    \begin{cases}
                        \phi + \displaystyle\int_{2\alpha}^{\delta_\alpha} (s - 2\alpha)\,\mathrm{d}s +
                        \displaystyle\int_{\delta_\alpha}^{1 + 2\alpha - \phi}
                        (\rho\cdot(1 + \alpha - s) + \epsilon) \,\mathrm{d}s
                        \quad &\text{ if } 1 + 2\alpha - \phi > \delta_\alpha\nonumber\\
                        \phi + \displaystyle\int_{2\alpha}^{1 + 2\alpha -\phi}(s - 2\alpha)\,\mathrm{d}s
                        &\text{ if } 1 + 2\alpha - \phi \leqslant \delta_\alpha \nonumber
                    \end{cases} \nonumber
            \end{eqnarray}
            That is (see Fig \vref{graph:EJ} for a plot)
            \begin{description}
                \item [if $\phi < 1 + 2\alpha - \delta_\alpha$:]
                    $$
                        \phi + E(\mathcal{J}^{\ast}(\phi_\leftarrow\hat{\phi}))
                        = \phi + \frac{1}{2}\cdot(\delta_\alpha - 2\alpha)^2 + \frac{1}{2}\,(1 + 2\alpha - \delta_\alpha - \phi)[2\epsilon + (1 - \delta_\alpha + \phi)\rho]
                    $$
                \item [if $\phi \geqslant 1 + 2\alpha - \delta_\alpha$:]
                    $$
                        \phi + E(\mathcal{J}^{\ast}(\phi_\leftarrow\hat{\phi}))
                        = \left(\frac{1 + \phi}{2}\right)^2
                    $$
            \end{description}
        \subsubsection{Absorption of $\alpha$-synchronized pairs}
            Suppose there is a pair of $\alpha$-synchronized oscillators A and B, whose states can be depicted in a standard
            form $(\phi_0, \alpha)$ where $\phi_0 = \alpha - d, 0 \leqslant d \leqslant \alpha$ (we can always make $\phi_0 > \alpha$ with
            symmetry of the bidirectional ring). Oscillator C ($\phi_1$) is B's right side neighbor. If $\mathbbm{D}(\phi_\text{C}, \phi_\text{B}) > \alpha$,
            we will find out that B and C are likely to achieve $\alpha$-synchrony,  which looks like C is absorbed into an
            $\alpha$-synchronized pair near it.

            Because A and B have already achieved $\alpha$-synchrony, according to Theorem \ref{th:restore}, this pair
            will restore their $\alpha$-synchrony everytime they reacts to each other's pulse. In this estimation, we ignore the small chance
            that A may catch up with B, so B only reacts to C's pulse. Follow the mathematical expectation based approach,
            we may estimate B and C's state as $(\alpha, \phi + E(\mathcal{J}^{\ast}(\phi_\leftarrow\hat{\phi})))$.
            After B's reacting to C's firing, the state becomes
            $(h(\phi + E(\mathcal{J}^{\ast}(\phi_\leftarrow\hat{\phi}))), \alpha)$. Adding an expectation of D's influence
            on C again, the state goes to
            $$
                (h(\phi + E(\mathcal{J}^{\ast}(\phi_\leftarrow\hat{\phi}))), \alpha + E(\mathcal{J}^{\ast}(\alpha_\leftarrow\hat{\phi})))
            $$
            which is equivalent to
            $$
                (h(\phi + E(\mathcal{J}^{\ast}(\phi_\leftarrow\hat{\phi})))- E(\mathcal{J}^{\ast}(\alpha_\leftarrow\hat{\phi})), \alpha )
            $$
            in the next round it becomes
            $$
                \Bigg(\alpha, h\bigg(h\left(\phi + E(\mathcal{J}^{\ast}(\phi_\leftarrow\hat{\phi}))\right)- E(\mathcal{J}^{\ast}(\alpha_\leftarrow\hat{\phi}))\bigg)\Bigg)
            $$
            Note that $\mathcal{J}^{\ast}(\phi\leftarrow\hat\phi)$ defined in (\ref{EJump})
            has a prerequisite of $\phi > 2\alpha$, so we have to
            recalculate:
            \begin{eqnarray*}
                & & E(\mathcal{J}^{\ast}(\alpha\leftarrow\hat\phi))\\
                &=& E(\mathcal{J}(\hat\phi))\\
                &=& \int_{2\alpha}^{1}\mathcal{J}(s)\,\mathrm{d}s\\
                &=& \int_{2\alpha}^{\delta_\alpha}(s - 2\alpha)\,\mathrm{d}s + \int_{\delta_\alpha}^{1}\left(\rho\cdot(1 + \alpha - s) + \epsilon\right)\,\mathrm{d}s\\
                &=& \frac{1}{2}\,(\delta_\alpha-2\alpha)^2 + (1-\delta_\alpha)\left[\rho\,(1+\alpha)+\epsilon\right] - \frac{1}{2}\,(1-\delta_\alpha)^2
            \end{eqnarray*}
            We define the \emph{probability firing map}:
            \begin{equation}
                \mathbbm{H}(\phi) = h\bigg(h\left(\phi + E(\mathcal{J}^{\ast}(\phi_\leftarrow\hat{\phi}))\right)- E(\mathcal{J}^{\ast}(\alpha_\leftarrow\hat{\phi}))\bigg)
                \label{BigH}
            \end{equation}
            Then similar to \cite{MS1990} and \cite{MM1996}, the synchronization of B and C can be studied by
            analyzing the iteration of $\mathbbm{H}$. That is, whether C will be absorbed to an $\alpha$-synchronized
            pair (A,B) depends on whether $\exists k \in \mathbbm{N}$ s.t. $\mathbbm{H}^{(k)}(\phi) = 0$, here
            $\mathbbm{H}^{(k)}$ is an iteration of $\mathbbm{H}$: $\mathbbm{H}^{(k)} = \mathbbm{H}(\mathbbm{H}^{(k - 1)})$
        \subsubsection{Explanation of asynchrony phenomena with iteration of $\mathbbm{H}$}
\begin{figure}[b]
                \includegraphics[width = 8cm]{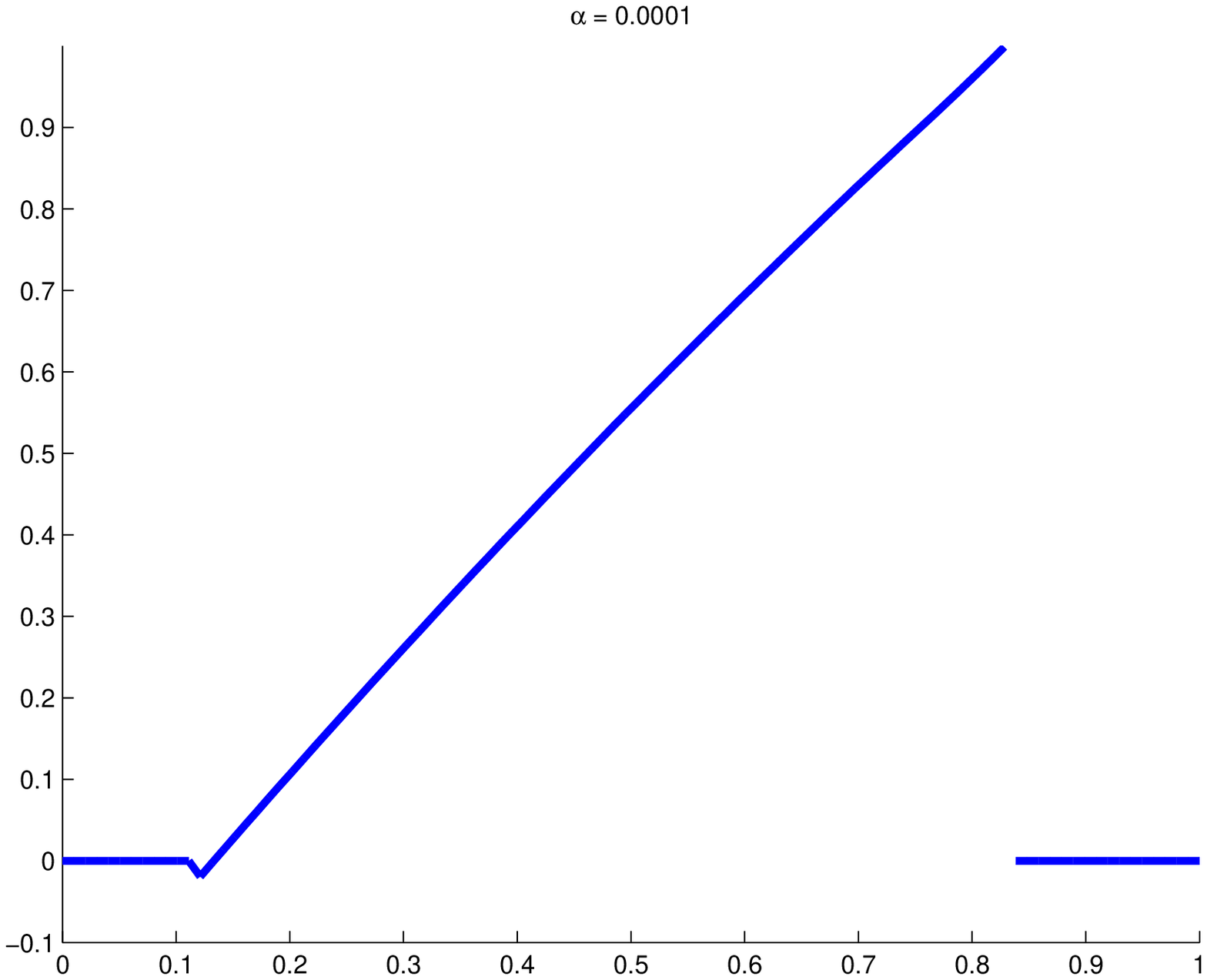}
                \includegraphics[width = 8cm]{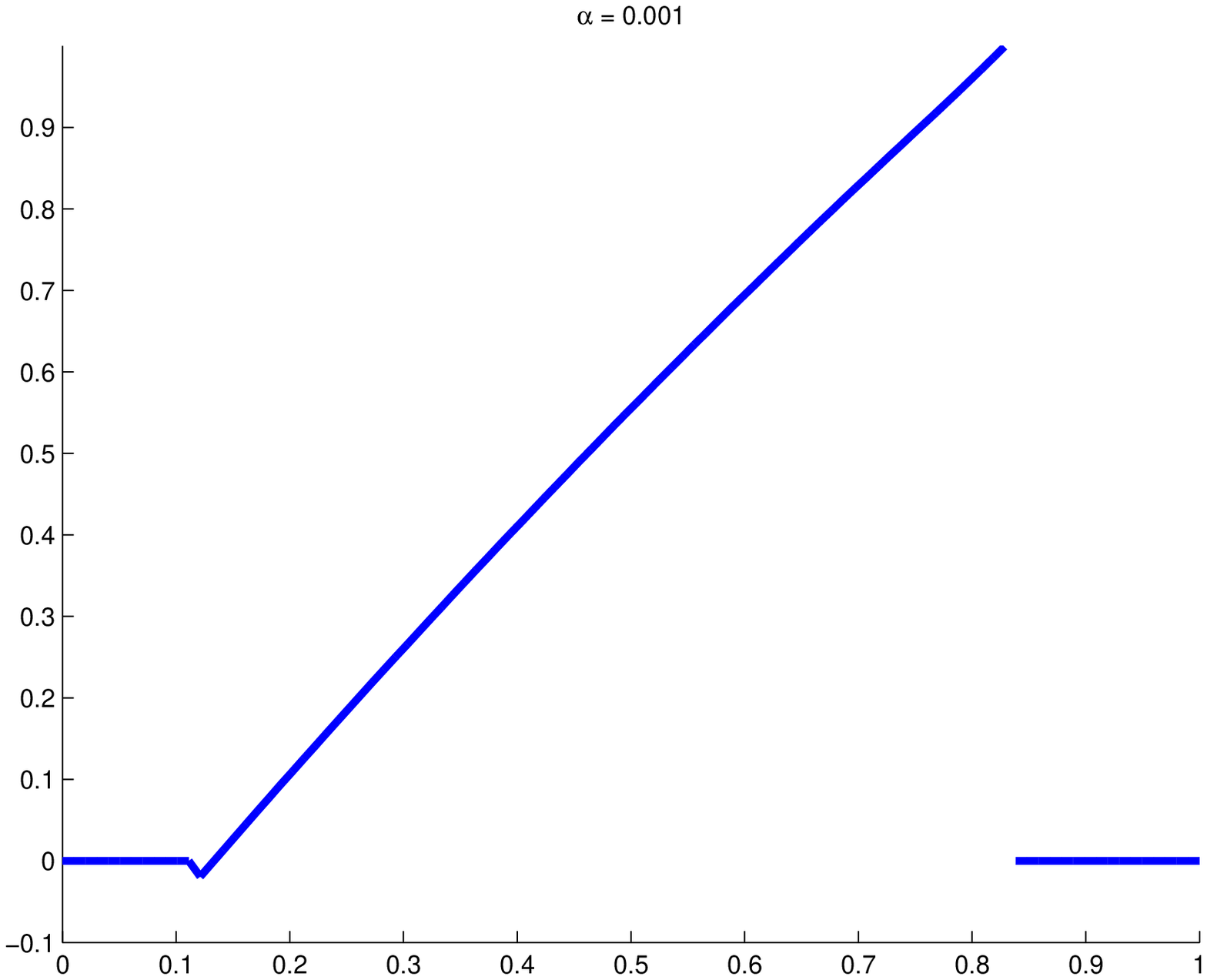}
                \includegraphics[width = 8cm]{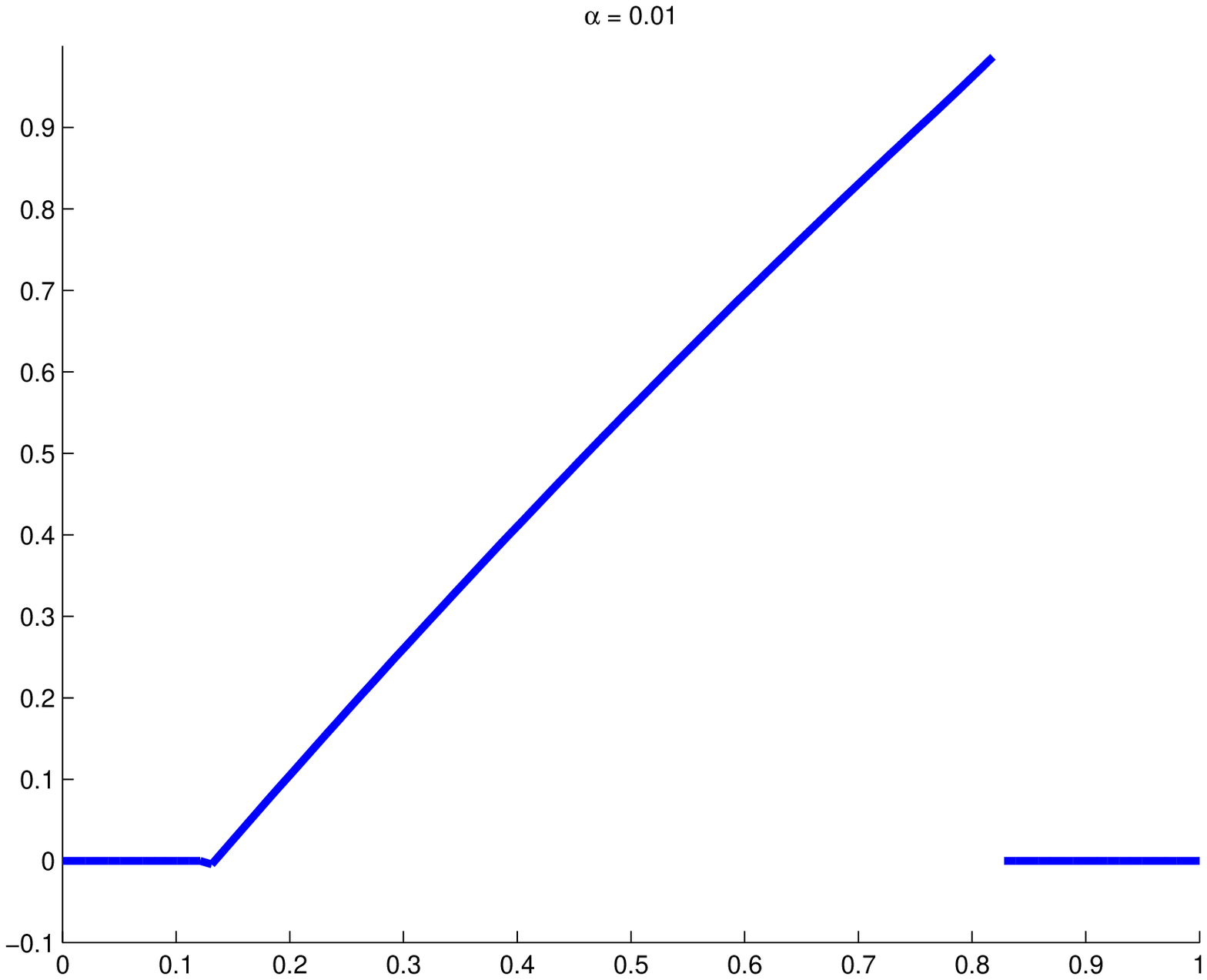}
                \includegraphics[width = 8cm]{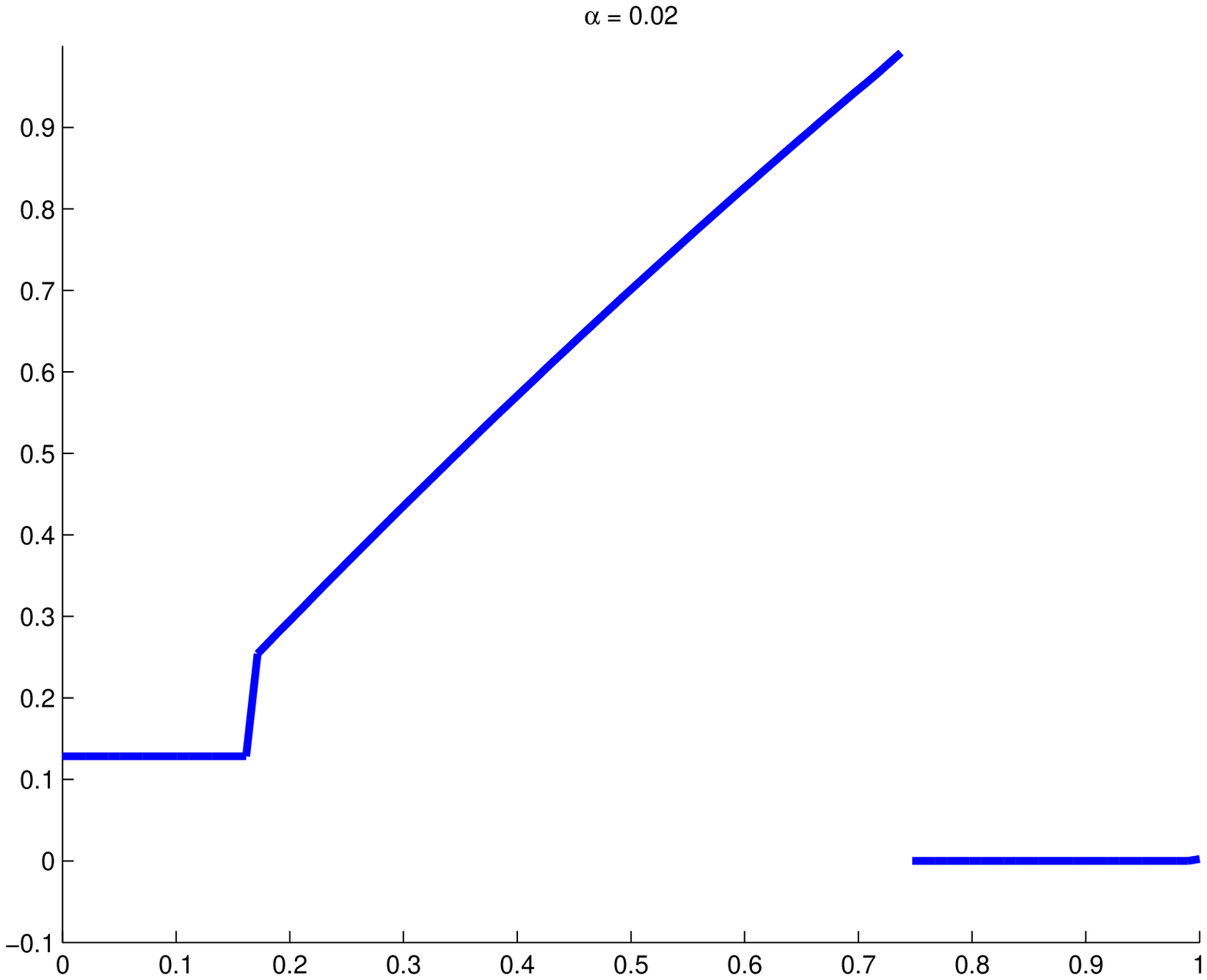}
                \includegraphics[width = 8cm]{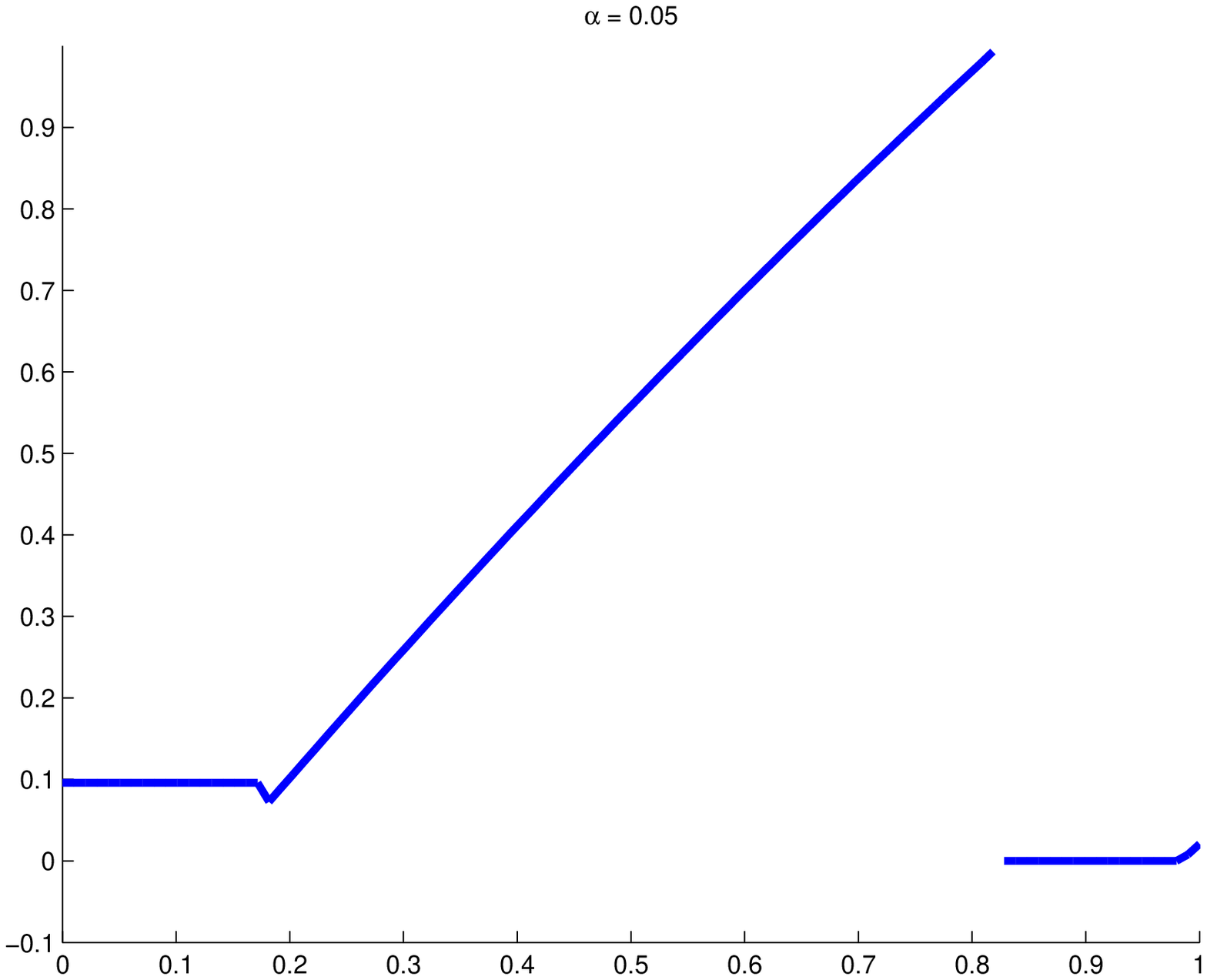}
                \includegraphics[width = 8cm]{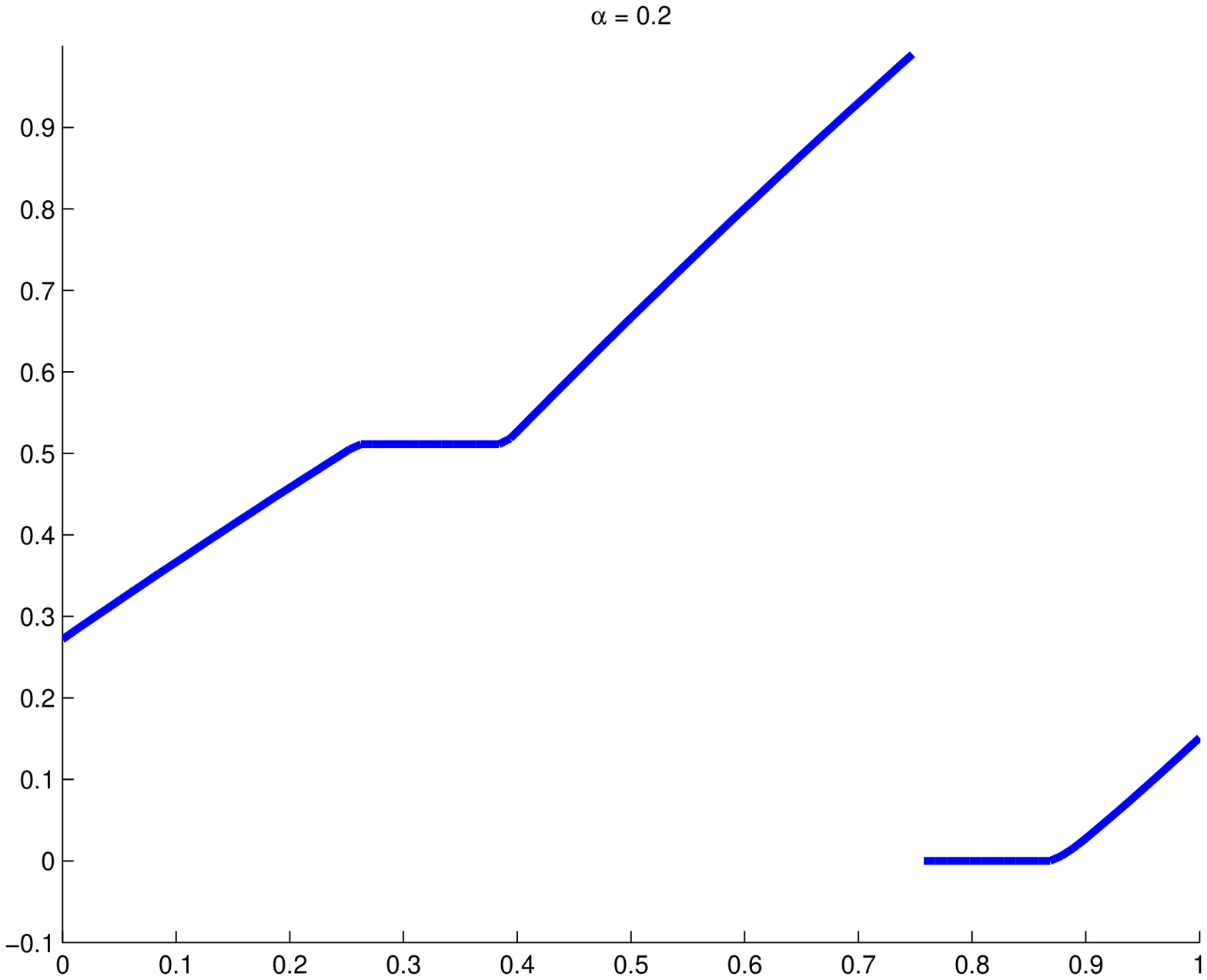}
                \caption{\label{graphBigH}Graph of $\mathbbm{H}(\phi)$ with differnet $\alpha$, when $\rho = 0.3$, $\epsilon = 0.01$}
            \end{figure}

            In purpose of analyzing the properties of $\mathbbm{H}$, its graphs are plotted with respect to different
            parameters. We noticed the change of the structure of $\mathbbm{H}$ 's graph with different $\alpha$.
            Also, in the experiments mentioned in previous section,
            we observed that with $\alpha \geqslant 0.1$, systems often fail to achieve $\alpha$-synchronization.
            With the graphs of $\mathbbm{H}$ with different $\alpha$, such phenomenons are easier to explain.
            From Fig. \vref{graphBigH}, we can see that with $\alpha < 0.1$, the $\mathbbm{H}$ 's graph is primarily
            consistied of a skew part and two flat parts, defined as $N_{\mathbbm{H}} = \{x | \mathbbm{H}(x) = 0\}$.
            Obviously, if $x \in N_{\mathbbm{H}}$, B and C will achieve synchrony in the next round. Even if
            $x \notin N_{\mathbbm{H}}$, as the slope of the skew part $[0, 1] \backslash N_{\mathbbm{H}}$ is greater
            than 1, the value of $\mathbbm{H}^{k}(x)$ will eventually ``fall'' in to $N_{\mathbbm{H}}$, which results in
            $\mathbbm{H}^{k + 1}(x) = 0$.

            As the value of $\alpha$ increases, the graph of $\mathbbm{H}(x)$ gradually becomes not so good looking.
            For $\alpha = 0.02$ in Fig \vref{graphBigH}, we can find ``pits'' on the graph, which prohibits the iteration
            to reach $\mathbbm{H}^{k}(x) = 0$. For example, the flat part $[0.1, 0.15]$ will cause every initial values
            that in it to reach a fixed point $x^\ast \approx 0.11$, i.e. $\forall x \in [0, 0.15]$,
            $\mathbbm{H}(x) = x^\ast$, where $\mathbbm{H}(x^\ast) = x^\ast \neq 0$. This is an explanation of
            the happening of phenonemons of asynchrony in $\alpha = 0.02$ observed in previous experiments in fig \vref{graph:asynchrony2}.

            Also on the graph of $\alpha = 0.05$, the same patterns are spotted:
            $\mathbbm{H}([0, 0.15]) \approx 0.1 \in [0, 0.15]$, which creates a fixed point of $\mathbbm{H}$ that attracts initial
            values in $[0, 0.15]$.

            However, with $\alpha = 0.2$, the graph of $\mathbbm{H}$ in Fig (\ref{graphBigH}) is in another pattern. A rough observation on this graph
             tells: for turning points $x_1 \approx 0.25$, $x_2 \approx 0.4$, $x_3 \approx 0.75$, $x_4 \approx 0.9$
              of $y = \mathbbm{H}(x)$,
            $\mathbbm{H}([0, x_1]) > x_1$, $\mathbbm{H}([x_1, x_2]) > x_2$, $\mathbbm{H}'([x_2, x_3]) > 1$
            and $\mathbbm{H}([x_3, x_4]) = 0$, $\mathbbm{H}([x_4, 1]) \in [0, x_1]$, which means
            \begin{eqnarray*}
                \mathbbm{H}&:&[x_4, 1] \Rrightarrow [0, x_1]\\
                \mathbbm{H}&:&[0, x_1] \Rrightarrow  [x_1, x_2]\\
                \mathbbm{H}&:&[x_1, x_2] \Rrightarrow [x_2, x_3]\\
                \mathbbm{H}&:&[x_2, x_3] \Rrightarrow\Rrightarrow 0
            \end{eqnarray*}
            So, for an arbitrary initial point
            $\phi_0 \in [0, 1]$, with iteration of $\phi_{k + 1} = \mathbbm{H}(\phi_k)$, we will inevitably have
            $\phi_n = 0$ for some $n \in \mathbb{N}$.
\newpage
\begin{minipage}[b]{10cm}
\section{Conclusions}
    In this paper we have proven that in each pair that once achieved $\alpha$-synchronization will
    restore their synchrony no matter how their neighbors acts.
    Then with comparison between the numerical simulation results and the properties of $\mathbbm{H}$,
    we have successfully established a cause-and-effect
    link between the synchronization behavior of the systems
    and their probability firing map $\mathbbm{H}$. With this application of $\mathbbm{H}$, it is reasonable to
    believe that the probability based approach presented in section (\ref{section:dynamics}) reaches the point of
    the problem.
\section{Further works}
    On Fig \vref{graph:rho-epsilon-t1}, a non-linear change of timecost with respect to $\alpha$ is clearly seen. But
    this phenomenon is not yet studied in this paper. Also, in section (\ref{section:dynamics}), we ignored
    a lot of affecting factors that may attribute to the result of the system's evolution. And what is more,
    the reason of forming of patterns in $\mathbbm{H}$ should be searched with analytic studies on
    $E(\mathcal{J}^\ast(\phi\leftarrow\hat\phi))$
    Besides, we would like to inspect synchronization phenomena when a central ``host'' oscillator is added,
    which has connection to all, or part of the oscillators on the circumference.
    We hope to work on those topics in further studies.
\section*{Acknowledgements}
    The author of this paper benefited a lot from Prof. Tianping Chen's seminar, School of Mathematical Sciences, Fudan University.
\end{minipage}
\hfill
\begin{minipage}[t]{4cm}
    \begin{center}
        \includegraphics[height = 21cm]{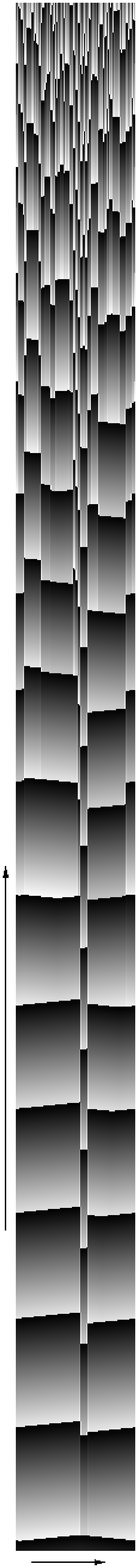}\\
    \end{center}
    {
        \small
        \renewcommand{\baselinestretch}{.3}
        \figcaption{
            This figure demonstrates an $\alpha$-synchronization process of 100 oscillators. $x$ axis represents
            the index of oscillators, $y$ axis represents time, grey level shows the phase of the oscillator
            (blackest: $\phi = 0$, whitest: $\phi = 1$). Inspired by \cite{VC1998}.
            \label{graph:greylevel}
        }
    }
\end{minipage}

\nocite{DP2003}
\nocite{BBH2004}
\bibliographystyle{unsrt}

\end{document}